# High-resolution spectroscopy of $Pr^{3+}$ ions in $YAl_3(BO_3)_4:Pr^{3+}$. Crystal-field, hyperfine, and electron-deformation interactions

B. Z. Malkin[1], T. A. Igolkina[2,3], K. N. Boldyrev[4], A. Baraldi[5], E. P. Chukalina[2], M. N. Popova[2,*]

[1] *Kazan Federal University, Kazan 420008, Russia*

[2] *Institute of Spectroscopy, Russian Academy of Sciences, Troitsk, Moscow 108840, Russia*

[3] *Moscow Institute of Physics and Technology (National Research University), Dolgoprudnyi, Moscow region 141700, Russia*

[4] *Beijing Institute of Technology (BIT), Zhuhai BIT, Zhuhai 519088, P. R. China*

[5] *Dept. of Mathematical, Physical and Computer Sciences, University of Parma, Parma 43124, Italy*

**Abstract**

Optical transmission spectra of $YAl_3(BO_3)_4$ crystals doped with the $Pr^{3+}$ ions in concentrations 1 and 2.5 at. % were studied by high-resolution (up to 0.05 $cm^{-1}$) Fourier spectroscopy, including in magnetic field parallel to the trigonal *c* axis of the crystal. The *g* factors of several crystal-field doublets of $Pr^{3+}$ were determined. The crystal-field calculations performed using the exchange-charge model and high-resolution spectroscopy data allowed us to obtain a physically reasonable set of crystal-field parameters. The observed splitting of a number of doublets in zero magnetic field is explained by the presence of random lattice deformations. Simulation of the profiles of observed deformational doublets was carried out taking into account both hyperfine and electron-deformation interactions. The width of the distribution function of random strains was estimated. The main sources of random strains in $YAl_3(BO_3)_4:Pr^{3+}$ crystals are discussed.

* corresponding author, e-mail: popova@isan.troitsk.ru

# I. INTRODUCTION

Yttrium aluminum borate YAl$_3$(BO$_3$)$_4$ (YAB) is a multifunctional material with attractive physical and chemical properties. It is characterized by mechanical, chemical and thermal stability [1], uniquely high thermal conductivity [2], transparency in a wide spectral range, and high nonlinear optical coefficient [3], [4]. Using a 2.94 mm thick YAB crystal, 240 kW peak power at 266 nm (the fourth harmonic of a Nd:YAG laser) was achieved [3]. Doped with rare-earth (RE) ions, YAB crystals are used as nonlinear active media for self-frequency-doubling and self-frequency-summing lasers (see, e.g., [5-8]). YAB crystallizes in the *R*32 noncentrosymmetric space symmetry group and belongs to the family of compounds with the structure of the natural mineral huntite [9]. RE$^{3+}$ ions substitute for Y$^{3+}$ ions and reside in isolated REO$_6$ distorted prisms with *D*$_3$ point symmetry group, which do not share oxygen atoms with other REO$_6$ units. Such structural feature results in a low concentration quenching of RE luminescence [10,11]; as a consequence, YAB crystals with high concentration of RE ions are used as active media for microchip lasers [12-15]. Concentrated REAl$_3$(BO$_3$)$_4$ crystals are multiferroics [16-19], in particular, a giant magnetoelectric effect was demonstrated on HoAl$_3$(BO$_3$)$_4$ [17]. It has been shown that magnetoelectric properties of REAl$_3$(BO$_3$)$_4$ crystals can be well described using crystal-field parameters obtained from spectroscopic data [19]. YAB:RE$^{3+}$ crystals are efficient phosphors [20-22], they can be used, in particular, for temperature sensing [23].

Luminesce of praseodymium ions Pr$^{3+}$ imbedded into a crystal or glass is effectively employed in modern optical technologies. Such crystals are used as phosphors [24-27] and laser media [28-32]. Ultra-sensitive low-temperature luminescent thermometry based on Boltzmann behavior of the crystal-field levels of Pr$^{3+}$ was recently demonstrated with CaNb$_2$O$_6$:Pr$^{3+}$ [33]. Pr$^{3+}$-doped crystals are considered for applications in quantum informatics devices [34-40]. The long coherence times of hyperfine structure levels required in this case have been recorded for Pr$^{3+}$ in oxide crystals Y$_2$SiO$_5$ [36-39] and La$_2$(WO$_4$)$_3$ [40].

The above justifies the interest in studying the crystal-field and hyperfine structure of the praseodymium levels in YAB:Pr$^{3+}$ crystal. A recent high-resolution study of the temperature-dependent polarized absorption spectra of YAB:Pr$^{3+}$ (1at. %) crystals [41] allowed us to specify the energies of CF levels, the previous information on which was incomplete or contradictory [42,43], and to reveal deformation splittings of spectral lines corresponding to optical transitions between singlet and doublet CF levels. Here, we add a new experimental information on magnetic *g* factors and deformation splittings for Pr$^{3+}$ doped into YAB crystals at concentrations of 1 at. % and 2.5 at. % and grown with different fluxes, and we simulate the spectral line shape based on CF calculations, taking into account both hyperfine and electron-deformation interactions.

## II. EXPERIMENTAL DETAILS

In addition to the YAB:$Pr^{3+}$(1 at. %) crystal grown at the L.V. Kirenskiii Institute of Physics in Krasnoyarsk using the solution-melt method with the $Bi_2Mo_3O_{12}$-based flux [44] and studied in [41], a YAB:$Pr^{3+}$(2.5 at. %) crystal grown at the Moscow State University by N.I. Leonyuk using the same method but with the $K_2Mo_3O_{10}$-based flux [9] was studied. Also, the deformation splittings in the absorption spectra of a YAB:$Pr^{3+}$(1 at. %) crystal grown with the $K_2O/MoO_3/B_2O_3$ mixed flux at the University of Parma [43] were analysed.

Transmission spectra in the range of 2000–23,000 $cm^{-1}$ were recorded on a Bruker IFS 125 HR Fourier spectrometer. Measurements were performed in the temperature range of 5–300 K using a Sumitomo RP-082E2S closed helium cycle cryostat. Temperature control and stabilization were performed using a Lake Shore Model 335 dual-channel temperature controller. Transmission spectra were recorded in unpolarized light (**k**⊥**c**), π (**k**⊥**c**, **E**∥**c**) and σ (**k**⊥**c**, **E**⊥**c**) polarizations. The spectral resolution was 0.05 $cm^{-1}$ when recording spectra in unpolarized light and 0.3 $cm^{-1}$ in π and σ polarizations. A KRS-5-based polarizer was used for the infrared region, and a film polarizer was used for the visible region.

To determine the $g$ factors, the splittings of spectral lines corresponding to singlet-doublet transitions in a magnetic field were measured. The magnetic field parallel to the crystal's $c$ axis was generated by permanent neodymium-iron-boron magnets placed in a cryostat next to the sample. The field strength, B = 0.595 ± 0.06 T, was measured using a Hall magnetometer.

## III. EXPERIMENTAL RESULTS

### A. Energies, symmetries, and g factors for the cryatal-field levels of $Pr^{3+}$ in YAB:$Pr^{3+}$

The wave functions of the CF levels of the $Pr^{3+}$ ion, occupying a position with point symmetry group $D_3$ in YAB, are transformed according to two non-degenerate irreducible representations $\Gamma_1$ and $\Gamma_2$ and one doubly degenerate $\Gamma_3$. The analysis of the temperature-dependent polarized absorption spectra of the YAB:$Pr^{3+}$(1 at. %) crystal, with the use of selection rules, allowed us to determine energies and symmetries (irreducible representations) of 43 CF levels of $Pr^{3+}$ in $YAl_3(BO_3)_4$:$Pr^{3+}$ [41]. These data are given in Table I together with the data for the isostructural compound $PrFe_3(BO_3)_4$.

TABLE I. Energies $E$ (cm$^{-1}$) of the CF levels of Pr$^{3+}$ in YAl$_3$(BO$_3$)$_4$:Pr$^{3+}$, experimentally found and calculated. Irreducible representations $\Gamma$ and calculated $g$ factors $g_\parallel$ are also presented. Last two columns show the experimental data on the isostructural PrFe$_3$(BO$_3$)$_4$.

| $^{2S+1}L_J$ | i | Exp. YAB-Pr [41] $E, \Gamma$ | Theory YAB-Pr This work $E$ | $\Gamma$ | $g_\parallel$ | PrFe$_3$(BO$_3$)$_4$ [45] $E$ | $\Gamma$ |
|---|---|---|---|---|---|---|---|
| $^3H_4$ | 1 | 0 $\Gamma_2$ | 0 | $\Gamma_2$ | 0 | 0 | $\Gamma_2$ |
|  | 2 | 23 $\Gamma_1$ | 23 | $\Gamma_1$ | 0 | 48.5 | $\Gamma_1$ |
|  | 3 | 226 $\Gamma_3$ | 225 | $\Gamma_3$ | 1.80 | 192 | $\Gamma_3$ |
|  | 4 | 255 | 248 | $\Gamma_3$ | 0.88 | 275 | $\Gamma_3$ |
|  | 5 | 493-635 | 493 | $\Gamma_3$ | 2.34 | 500 | $\Gamma_3$ |
|  | 6 | 493-635 | 531 | $\Gamma_1$ | 0 | 560 | $\Gamma_1$ |
| $^3H_5$ | A | 2196 $\Gamma_3$ | 2189 | $\Gamma_3$ | 7.481 | 2182.3 | $\Gamma_3$ |
|  | B | 2272 | 2274 | $\Gamma_2$ | 0 | 2275.2 | $\Gamma_1$ |
|  | C | - | 2278 | $\Gamma_1$ | 0 | 2295.5 | $\Gamma_2$ |
|  | D | - | 2329 | $\Gamma_3$ | 7.45 | 2306.6 | $\Gamma_3$ |
|  | E | - | 2491 | $\Gamma_3$ | 6.26 | -- | $\Gamma_3$ |
|  | F | - | 2551 | $\Gamma_3$ | 1.22 | 2588 | $\Gamma_3$ |
|  | G | - | 2613 | $\Gamma_2$ | 0 | -- | $\Gamma_2$ |
| $^3H_6$ | A | 4295.5 $\Gamma_3$ | 4307 | $\Gamma_3$ | 10.48 | 4292 | $\Gamma_3$ |
|  | B | 4338.7 $\Gamma_3$ | 4364 | $\Gamma_3$ | 8.33 | 4355.2 | $\Gamma_3$ |
|  | C | - | 4491 | $\Gamma_2$ | 0 | -- | $\Gamma_2$ |
|  | D | - | 4511 | $\Gamma_1$ | 0 | 4503.5 | $\Gamma_1$ |
|  | E | - | 4662 | $\Gamma_2$ | 0 | -- | $\Gamma_2$ |
|  | F | 4707 $\Gamma_1$ | 4673 | $\Gamma_1$ | 0 | 4641 | $\Gamma_1$ |
|  | G | 4727.8 $\Gamma_3$ | 4693 | $\Gamma_3$ | 3.19 | 4748 | $\Gamma_3$ |
|  | H | 4817.8 $\Gamma_3$ | 4788 | $\Gamma_3$ | 1.60 | 4819.3 | $\Gamma_3$ |
|  | I | 4845 $\Gamma_1$ | 4808 | $\Gamma_1$ | 0 | 4850 | $\Gamma_1$ |
| $^3F_2$ | A | 5103 $\Gamma_3$ | 5111 | $\Gamma_3$ | 0.08 | 5079 | $\Gamma_3$ |
|  | B | 5187 $\Gamma_1$ | 5147 | $\Gamma_1$ | 0 | 5141.5 | $\Gamma_1$ |
|  | C | 5206 $\Gamma_3$ | 5188 | $\Gamma_3$ | 2.57 | 5190.6 | $\Gamma_3$ |
| $^3F_3$ | A | 6484.3 $\Gamma_2$ | 6492 | $\Gamma_3$ | 0.72 | 6471.1 | $\Gamma_3$ |
|  | B | 6498.6 $\Gamma_3$ | 6497 | $\Gamma_2$ | 0 | 6482.5 | $\Gamma_2$ |
|  | C | 6524 | 6534 | $\Gamma_3$ | 3.60 | 6508 | $\Gamma_3$ |
|  | D | 6583.8 $\Gamma_1$ | 6566 | $\Gamma_2$ | 0 | 6555 | $\Gamma_2$ |
|  | E | 6589.8 $\Gamma_2$ | 6581 | $\Gamma_1$ | 0 | 6571.7 | $\Gamma_1$ |
| $^3F_4$ | A | 6855.5 $\Gamma_1$ | 6852 | $\Gamma_1$ | 0 | 6863 | $\Gamma_1$ |
|  | B | 6891.3 $\Gamma_3$ | 6888 | $\Gamma_3$ | 1.64 | 6885.8 | $\Gamma_3$ |
|  | C | 6930 $\Gamma_2$ | 6936 | $\Gamma_2$ | 0 | 6957.3 | $\Gamma_2$ |
|  | D | 6965 $\Gamma_3$ | 6972 | $\Gamma_3$ | 3.47 | 6957.5 | $\Gamma_3$ |
|  | E | - | 7123 | $\Gamma_1$ | 0 | 7099.5 | $\Gamma_1$ |

|  | F | 7150 | 7138 | $\Gamma_3$ | 6.59 | 7107.8 | $\Gamma_3$ |
|---|---|---|---|---|---|---|---|
| $^1G_4$ | A | 9707.4 | 9706 | $\Gamma_1$ | 0 | 9728.5 | $\Gamma_1$ |
|  | B | - | 9774 | $\Gamma_3$ | 0.94 | 9776.7 | $\Gamma_3$ |
|  | C | 9909 $\Gamma_2$ | 9909 | $\Gamma_2$ | 0 | 9925.6 | $\Gamma_3$ |
|  | D | 10178 | 9946 | $\Gamma_3$ | 3.16 | 9946 | $\Gamma_2$ |
|  | E | 10216 | 10261 | $\Gamma_3$ | 5.63 | 10113.8 | $\Gamma_3$ |
|  | F | - | 10289 | $\Gamma_1$ | 0 | 10193.4 | $\Gamma_1$ |
| $^1D_2$ | A | 16512 $\Gamma_1$ | 16571 | $\Gamma_1$ | 0 | 16491 | $\Gamma_1$ |
|  | B | 16754 $\Gamma_3$ | 16798 | $\Gamma_3$ | 1.81 | 16791 | $\Gamma_3$ |
|  | C | 17140 $\Gamma_3$ | 17088 | $\Gamma_3$ | 0.20 | 17080 | $\Gamma_3$ |
| $^3P_0$ | A | 20603 $\Gamma_1$ | 20596 | $\Gamma_1$ | 0 | 20567 | $\Gamma_1$ |
| $^3P_1 +$ | A | 21003 $\Gamma_2$ | 20952 | $\Gamma_2$ | 0 | 20848 | $\Gamma_2$ |
| $^1I_6$ | B | 21014 $\Gamma_1$ | 20959 | $\Gamma_1$ | 0 | 20856 | $\Gamma_1$ |
|  | C | 21163 $\Gamma_3$ | 21171 | $\Gamma_3$ | 3.07 | 21100 | $\Gamma_3$ |
|  | D | - | 21282 | $\Gamma_2$ | 0 | 21280 | $\Gamma_2$ |
|  | E | 21330 $\Gamma_1$ | 21346 | $\Gamma_1$ | 0 | 21382 | $\Gamma_3$ |
|  | F | 21353 $\Gamma_3$ | 21363 | $\Gamma_3$ | 0.02 | 21420 | $\Gamma_1$ |
|  | G | - | 21409 | $\Gamma_3$ | 0.18 | 21453 | $\Gamma_3$ |
|  | H | - | 21704 | $\Gamma_2$ | 0 | -- | $\Gamma_2$ |
|  | I | - | 21745 | $\Gamma_3$ | 4.96 | -- | $\Gamma_3$ |
|  | K | 21816 $\Gamma_1$ | 21764 | $\Gamma_1$ | 0 | 21760 | $\Gamma_1$ |
|  | L | 21858.6 $\Gamma_3$ | 21819 | $\Gamma_3$ | 6.43 | 21810 | $\Gamma_3$ |
| $^3P_2$ | A | 22261 $\Gamma_1$ | 22260 | $\Gamma_1$ | 0 | 22193 | $\Gamma_1$ |
|  | B | 22367 $\Gamma_3$ | 22358 | $\Gamma_3$ | 0.96 | 22334 | $\Gamma_3$ |
|  | C | 22450 $\Gamma_3$ | 22442 | $\Gamma_3$ | 4.87 | -- | $\Gamma_3$ |
| $^1S_0$ | A | - | 45968 | $\Gamma_1$ | 0 | - | - |

In order to obtain additional information on the wave functions of the $\Gamma_3$ doublets, the transmission spectra of the YAB:Pr$^{3+}$(1 at. %) sample were measured in a constant magnetic field $B = 0.595 \pm 0.06$ T, parallel to the $c$ axis, at a temperature of 5 K. It was possible to measure the Zeeman splittings $\Delta E$ only for the three narrowest lines of the singlet-doublet transitions from the ground singlet to the doublets. They are listed in Table II together with $g$ factors calculated as $g_\parallel = \Delta E/\mu_B B$. In the last column of Table II, the values calculated using wave functions obtained from CF calculations are presented

TABLE II. Measured splittings $\Delta E$ (cm$^{-1}$) of singlet-doublet optical transitions in magnetic field **B**||**c** and corresponding $g$ factors $g_\parallel = \Delta E/\mu_B B$. The last column gives the calculated values of $g_\parallel$.

| Transition | $\Delta E, cm^{-1}$ | $g_\parallel$ | $g_\parallel$, theory |
|---|---|---|---|
| $1\Gamma_2\,(^3H_4) \Rightarrow A\Gamma_3(^3H_5)$ | 2.2 ± 0.1 | 7.92 ± 1.15 | 7.48 |
| $1\Gamma_2\,(^3H_4) \Rightarrow A\Gamma_3(^3H_6)$ | 2.8 ± 0.1 | 10.1 ± 1.37 | 10.48 |
| $1\Gamma_2\,(^3H_4) \Rightarrow B\Gamma_3(^3H_6)$ | 2.5 ± 0.1 | 9 ± 1.26 | 8.335 |

### B. Doublet structure of singlet-doublet transition lines

Figure 1 shows the absorption line corresponding to the singlet-doublet transition $1\Gamma_2(^3H_4) \to B\Gamma_3(^3H_6)$ in YAB:Pr$^{3+}$ crystals with two different concentrations of praseodymium ions. The doublet structure is clearly visible, and the distance between the doublet components is noticeably greater in the spectrum of the crystal with a higher praseodymium concentration.

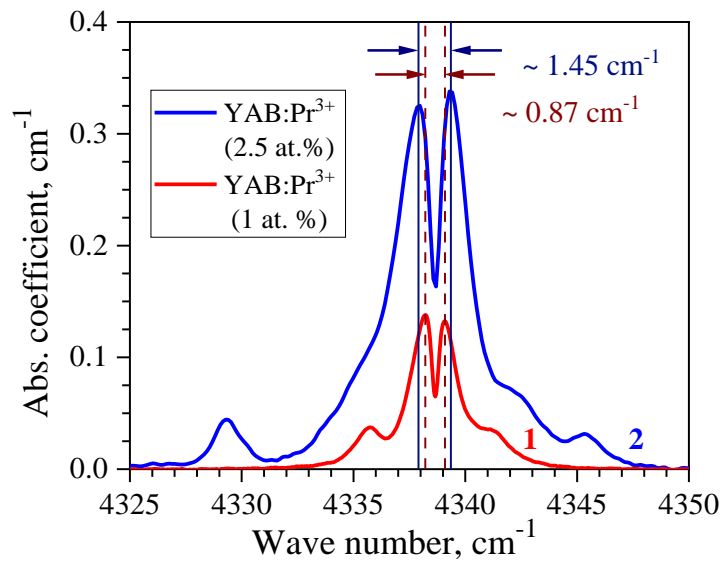

FIG. 1. Spectral absorption line of σ-polarized light corresponding to the transition $1\Gamma_2(^3H_4) \to B\Gamma_3(^3H_6)$ at T=5 K in YAB:Pr$^{3+}$ crystals with a praseodymium ion concentration of 1 at.% (red curve 1) and 2.5% (blue curve 2).

Praseodymium is a monoisotopic element with one stable isotope, $^{141}$Pr, with a nuclear spin of $I = 5/2$. As a result of hyperfine interactions, the non-Kramers $\Gamma_3$ doublets are split into six hyperfine components. No hyperfine structure was detected in the recorded spectra. However, the a doublet structure of some spectral lines related to the singlet-doublet transitions $1\Gamma_2(^3H_4) \to \Gamma_3$ was observed (see Figs. 1 and 2). Such characteristic doublet line shape indicates the presence of random low-symmetry local perturbations of the crystal field, which can be caused by point defects of the crystal lattice [46].

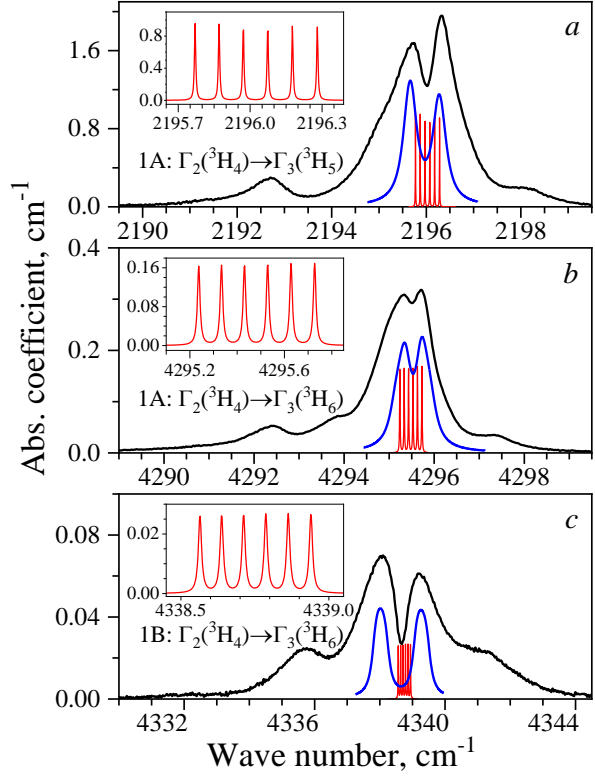

FIG. 2. Spectral absorption lines of unpolarized light related to the $\Gamma_2 \rightarrow \Gamma_3$ transitions in the YAB:Pr$^{3+}$(1 at. %) crystal at $T$=5 K. The black line is the experiment, the red line is the calculated six-component hyperfine structure, the blue line is the calculated envelope of the hyperfine transitions from the ground singlet in the field of random lattice deformations under the assumption of a Lorentzian line shape of individual transitions with a half-width of 0.053 cm$^{-1}$.

Table III compares the observed splittings in the spectra of three different YAB:Pr$^{3+}$ crystals. In crystals with 1 at. % praseodymium concentrations, grown with very different fluxes, the deformation splittings $\Delta(1\%)$ are identical, whereas in YAB:Pr$^{3+}$(2.5 at. %) this splitting $\Delta(2.5\%)$ is significantly larger. The line intensity ratio confirms that the concentration ratio of praseodymium in these crystals is 2.5. (see Fig. 1). At the same time, the ratio of the observed splittings is noticeably smaller than 2.5. In [47,48], it was shown that during crystal growth using the solution-melt method, flux components enter the crystal. Mo$^{3+}$ ions replace Al$^{3+}$ ions, and Bi$^{3+}$ takes the place of Y$^{3+}$, creating additional defects [47]. Apparently, these defects as well as some other intrinsic defects of the crystal (e.g., vacancies), contribute to the deformation-induced splitting of RE$^{3+}$ spectral lines in YAB:RE$^{3+}$ crystals. At low concentration of defects, it is natural to assume an additivity of contributions from different kinds of defects and proportionality of each contribution to the concentration and the defect strength [46] and to present the deformation splittings as $\Delta(x\%) = \beta x + \Delta_i$. Here, $\beta$ is a coefficient proportional to the defect strength $\Omega_{Pr} \sim [r(\text{Pr}^{3+}) - r(\text{Y}^{3+})]$ and $\Delta_i$ is the contribution from intrinsic defects. Assuming that $\Delta_i$ does not depend on the flux used (which was confirmed for the two studied YAB:Pr$^{3+}$(1 at.%) crystals) and using the experimental data from Table III for the deformation splitting of the B$\Gamma_3$($^3$H$_6$) doublet in crystals with a praseodymium concentration of 1 at.% and 2.5% (see Fig. 1), we obtain $\Delta_i = 0.48\pm0.07$ cm$^{-1}$ for the contribution to the splitting caused by intrinsic defects.

TABLE III. Observed deformation splittings $\Delta$ of several lines of singlet-doublet transitions in different YAB:Pr crystals grown with different fluxes (indicated in the Table).

| Optical transition | Frequency (cm$^{-1}$) | $\Delta$ (cm$^{-1}$) | | |
|---|---|---|---|---|
| | | YAB:Pr(1%) Bi$_2$Mo$_3$O$_{12}$ | YAB:Pr(2.5%) K$_2$Mo$_3$O$_{10}$ | YAB:Pr(1%) K$_2$O/MoO$_3$/B$_2$O$_3$ |
| $1\Gamma_2(^3H_4) \to A\Gamma_3(^3H_5)$ | 2196 | 0.52±0.03 | 0.71±0.03 | 0.55±0.02 |
| $1\Gamma_2(^3H_4) \to A\Gamma_3(^3H_6)$ | 4295.5 | 0.34±0.03 | 0.53±0.03 | 0.39±0.04 |
| $1\Gamma_2(^3H_4) \to B\Gamma_3(^3H_6)$ | 4338.7 | 0.87±0.03 | 1.45±0.03 | 0.94±0.05 |

## IV. SIMULATION OF THE ABSORPTION SPECTRA
### A. Crystal-field parameters for the Pr$^{3+}$ ion in YAl$_3$(BO$_3$)$_4$:Pr$^{3+}$

The simulation of the absorption spectra of the YAl$_3$(BO$_3$)$_4$ crystal activated by Pr$^{3+}$ ions, which replace Y$^{3+}$ ions at the sites of the crystal lattice with local symmetry $D_3$, was performed on the basis of calculations of the energy levels and the corresponding wave functions of the impurity ion by the method of successive numerical diagonalization of the Hamiltonian defined in the full space of electron-nuclear states of the ground electron configuration $4f^2$ with the dimension $D=C_{14}^2(2I+1)=546$ ($I=5/2$ is the nuclear spin of the only stable isotope of praseodymium $^{141}$Pr),

$$H = H_0 + H_{CF} + H_Z + H_{HFM} + H_{HFQ} + H_{el\text{-}def.} \quad (1)$$

In operator (1), the first term,

$$H_0 = \zeta \sum_j \mathbf{l}_j \mathbf{s}_j + \alpha \hat{L}^2 + \beta \hat{G}(G_2) + \gamma \hat{G}(R_7) + \sum_q (F^q \hat{f}_q + P^q \hat{p}_q + M^q \hat{m}_q), \quad (2)$$

presented in standard form [49], includes the energies of electrostatic interaction between $4f$ electrons, spin-orbit interaction ($\mathbf{l}_j$ and $\mathbf{s}_j$ are the orbital and spin moments of the j-th electron, respectively, j=1.2), additional contributions due to mixing of the ground and excited configurations, and relativistic interactions. The second term in (1) defines the energy of $4f$ electrons in the crystal field:

$$H_{CF} = \sum_{j=1,2} \{B_0^2 C_0^{(2)}(j) + B_0^4 C_0^{(4)}(j) + B_0^6 C_0^{(6)}(j) + iB_{-3}^4[C_{-3}^{(4)}(j) + C_3^{(4)}(j)]$$

$$+ iB_{-3}^6[C_{-3}^{(6)}(j) + C_3^{(6)}(j)] + B_6^6[C_{-6}^{(6)}(j) + C_6^{(6)}(j)]\}. \quad (3)$$

Here $C_q^{(p)}$ are spherical tensor operators of rank $p$, written in a Cartesian coordinate system with the $z$ axis along the trigonal axis of symmetry $c$ and the $x$ axis along one of the $C_2$ axes in the $ab$ plane; $B_p^q$ are the crystal-field parameters calculated within the framework of the semi-phenomenological exchange-charge model [50]. This model takes into account the effects caused by the overlap of the wave functions of 4$f$ electrons with the wave functions of electrons localised on the nearest lattice ions. Summing the contributions of the electrostatic fields of point charges of ions ($B_{q,el}^p$) and exchange charges on Pr$^{3+}$ - O$^{2-}$ bonds ($B_{q,ex}^p$), we obtain

$$B_q^{p;} = B_{q,el}^p + B_{q,ex}^p, \tag{4}$$

$$B_{q,el}^p = -K_p^q \left(1 - \sigma_p\right) e^2 \langle r^p \rangle \sum_{L,s} \frac{q_s}{R_{Ls}} V_p^q (\theta_{Ls}, \varphi_{Ls}), \tag{5}$$

$$B_{q,ex}^p = \frac{2(2p+1)}{7} K_p^q e^2 \sum_v [G_s S_s^2(R_v) + G_\sigma S_\sigma^2(R_v) + \gamma_p G_\pi S_\pi^2(R_v)] \frac{V_p^q(\theta_v, \varphi_v)}{R_v}, \tag{6}$$

where $K_p^q$ are numerical coefficients [50], $\sigma_p$ are screening factors [51], $e$ is electron charge, $\langle r^p \rangle$ are moments of order $p$ of the radial-distribution density of 4$f$ electrons [52], (- $eq_s$) is the charge of an s-type ion in the unit cell $L$ of a crystal lattice with spherical coordinates $R_{Ls}, \theta_{Ls}$ and $\varphi_{Ls}$ (the origin of the coordinate system is located on the RE ion under consideration). The functions of the angular coordinates of the lattice ions $V_p^q$ are linear combinations of spherical functions [50]. In expression (6), the summation is carried out over the nearest neighbors of the RE ion, i.e., ligands $v$ at distances $R_v$ < 0.3 nm. The overlap integrals of 4$f$ wave functions of the RE ion |n=4,l,m⟩ (l=3, m=l_z, the quantization axis $z$ is directed along the vector $\boldsymbol{R_v}$) with the wave functions of electrons on the filled $ns$ ($l$=0), $np$ ($l$=1) shells of oxygen ions $S_s(R_v) = \langle 4,3,0|n,0,0 \rangle$, $S_\sigma(R_v) = \langle 4,3,0|n,1,0 \rangle$, $S_\pi(R_v) = \langle 4,3,1|n,1,1 \rangle$ are calculated using the corresponding radial Hartree-Fock wave functions presented in the literature [52,53]; $\gamma_2$=3/2, $\gamma_4$=1/3, $\gamma_6$=-3/2 [50], $G_s$, $G_\sigma$ and $G_\pi$ are variable parameters of the model, determined from the analysis of experimental data on CF levels of the RE ion.

The interaction energy of 4$f$ electrons with an external magnetic field **B** is determined by the operator $H_Z = -\boldsymbol{M} \cdot \boldsymbol{B}$, where $\boldsymbol{M} = -\mu_B(k\boldsymbol{L} + 2\boldsymbol{S})$ is the operator of the magnetic moment of the RE ion, $\mu_B$ is the Bohr magneton, and $k$ is the reduction factor of the orbital moment.

The operators of hyperfine magnetic dipole ($H_{HFM}$) and electric quadrupole ($H_{HFQ}$) interactions have the following form [54]:

$$H_{HFM} = A_{HF} \sum_j \{ \boldsymbol{I} \boldsymbol{l}_j + \frac{1}{2}[O_{2,j}^0(3s_{z,j}I_z - \boldsymbol{s}_j\boldsymbol{I}) + 3O_{2,j}^2(s_{x,j}I_x - s_{y,j}I_y)$$
$$+3O_{2,j}^{-2}(s_{x,j}I_y + s_{y,j}I_x) + 6O_{2,j}^1(s_{x,j}I_z + s_{z,j}I_x) + 6O_{2,j}^{-1}(s_{z,j}I_y + s_{y,j}I_z)]\}, \tag{7}$$

$$H_{\text{HFQ}} = \frac{e^2 Q(1-\gamma_\infty)}{4I(2I-1)} \sum_L q_L \frac{3z_L^2 - r_L^2}{r_L^5} I_0$$

$$-\frac{e^2 Q(1-R_Q)}{4I(2I-1)} \left\langle \frac{1}{r^3} \right\rangle_{4f} \sum_j [O_{2,j}^0 I_0 + 3O_{2,j}^2 I_2 + 3O_{2,j}^{-2} I_{-2} + 6O_{2,j}^1 I_1 + 6O_{2,j}^{-1} I_{-1}]. \quad (8)$$

Here, $O_p^q$ are linear combinations of spherical tensor operators $C_q^{(p)}$, which coincide with the Stevens operators in the space of eigenfunctions of the angular momentum operator [55], $A_{\text{HF}} = 2\mu_B \gamma 3_{4f_{Pr}}$ is the magnetic hyperfine interaction constant, $\gamma_{Pr} = 2\pi \cdot 13.05$ MHz/T is the nuclear gyromagnetic ratio, $Q = -5.9 \cdot 10^{-30}$ m$^2$ is the quadrupole moment of the nucleus, $R_Q = 0.1$ and $\gamma_\infty = -80$ are the Sternheimer screening and anti-screening factors [56,57], and $\langle r^{-3} \rangle_{4f} = 5$ at. u. [58]. The matrices of operators $\sum_j O_{2,j}^q s_{\alpha,j}$ were constructed in the full basis of Slater determinants of the electron configuration $4f^2$. The first term in (8) determines the interaction of the quadrupole moment of the nucleus with the gradient of the electric field of the ionic lattice in the coordinate system centered on the praseodymium ion under consideration ($\mathbf{r}_L$ is the radius vector of the ion with charge $eq_L$), the second term corresponds to the interaction with the electron shell, $I_0 = 3I_z^2 - I(I+1)$, $I_2 = I_x^2 - I_y^2$, $I_{-2} = I_x I_y + I_y I_x$, $I_1 = I_x I_z + I_z I_x$, $I_{-1} = I_y I_z + I_z I_y$.

The parameters of the Hamilton operator of a free ion (2) and the initial values of the CF parameters (see columns 1, 2, 4 in Table IV) were obtained from calculations of the energies of CF levels of Pr$^{3+}$ ions with varying charges of ions in the yttrium, aluminum, boron, and oxygen sublattices and the parameters $G_s, G_\sigma, G_\pi$ of the exchange-charge model to minimize deviations of the calculated energies from the measured ones. Results of variation: $q_Y = 3$, $q_{Al} = 2.68$, $q_B = 1.815$, $q_{O1}(9e) = -1.15$, $q_{O2}(9e) = -1.9$, $q_{O3}(18f) = -1.525$, $G_s = 2$, $G_\sigma = 2$ and $G_\pi = 1$; differences in the effective charges of oxygen ions in the Wyckoff positions 9e and 18f are due to covalent bonds in the BO$_3$ complexes. The final values of the CF parameters (column 5 in Table IV) were obtained by an additional procedure of directly varying the $B_q^p$ parameters, taking into account the local lattice deformation when replacing Y$^{3+}$ ions with an ionic radius of 0.90 Å with Pr$^{3+}$ ions with an ionic radius of 0.99 Å [59]. The energies and symmetries of CF levels and $g$ factors of the $\Gamma_3$ doublets calculated using this set of CF parameters are given in Table I. The calculated $g$ factors $g_\parallel = 2 < +|M_z|+>/\mu_B$ of the doublets ($|\pm\rangle$ are the doublet wave functions), the Zeeman splittings of which were observed in the transmission spectra in an external magnetic field, are in satisfactory agreement with the measurement data (see Table II).

TABLE IV. Parameters of operators (2) and (3) (cm$^{-1}$).

| Free-ion parameters | | Crystal-field parameters | | | | |
|---|---|---|---|---|---|---|
| Pr$^{3+}$:YAl$_3$(BO$_3$)$_4$ | | Pr$^{3+}$:YAl$_3$(BO$_3$)$_4$ | | | | PrFe$_3$(BO$_3$)$_4$ |
| 1 | 2 | 3 | 4 | 5 | 6 [43] | 7 [45] |
| $\zeta$ =755.7 $F^2$=67778 $F^4$=49603 $F^6$=32313 $\alpha$ =21.1 $\beta$ = -665 $\gamma$ =1634 | $P^2$=275 $P^4$=206 $P^6$=138 $M^0$=1.76 $M^2$=0.986 $M^4$=0.546 | $B_0^2$= $B_0^4$= $B_0^6$= $B_{-3}^4$= $B_{-3}^6$= $B_6^6$= | 433 -1784 540 1065 127 287 | 372 -1436 514 925 195 377 | 548±33 165±70 362±92 1336±49 457±67 -243±81 | 556 -1447 534 867 165 376 |

The CF parameters obtained in the present work (column 5 in Table IV) differ significantly from the corresponding parameters presented earlier in the literature (column 6 in Table IV), but are consistent with the parameters obtained in [45] (column 7 in Table IV) for the isostructural magnetically concentrated crystal PrFe$_3$(BO$_3$)$_{4CF}$. The agreement between the calculated energies of the CF levels and the experimental data has been noticeably improved. A number of contradictions between the arrangement of levels of different symmetries close in energy and a significantly underestimated total splitting of the $^1$D$_2$ multiplet are due to neglect of mixing the ground electron configuration 4$f^2$ with excited ones (4$f$6$p$, 4$f$6$s$, see [45]) in simulating the optical spectrum.

### B. Profiles of singlet-doublet transition lines

The Hamiltonian of the electron-deformation interaction has the following form:

$$H_{\text{el-def}} = \sum_{\Gamma=A,E} \sum_{k\lambda} V(\Gamma^k, \lambda) \, e(\Gamma^k, \lambda), \quad (9)$$

where the operators $V(\Gamma^k, \lambda)$ are defined by 30 independent parameters $b_p^q(\Gamma^k, \lambda)$,

$$V(\Gamma^k, \lambda) = \sum_{p,q} b_p^q(\Gamma^k, \lambda) O_p^q, \quad (10)$$

and $e(\Gamma^k, \lambda)$ are linear combinations of deformation tensor components $e_{\alpha\beta}$ (in the Cartesian coordinate system introduced above), transformed according to the irreducible representations $A$ and $E$ of the symmetry group $D_3$,

$$e(A^1) = (e_{xx} + e_{yy} + e_{zz})/\sqrt{3}, \quad e(A^2) = (2e_{zz} - e_{xx} - e_{yy})/\sqrt{12}, \quad (11)$$

$$e(E^1,1)=e_{xy}, \quad e(E^1,2)=(e_{xx} - e_{yy})/2, \quad e(E^2,1)=e_{xz}, \quad e(E^2,2)=e_{yz}, \quad (12)$$

The values of the parameters $b_p^q(\Gamma^k, \lambda)$ calculated within the framework of the exchange charge model [50] are given in Table V. The parameters of interaction with deformations $E^k$,1 and $E^k$,2 are related by the relations $b_p^q(E^k, 1) = b_p^{-q}(E^k, 2)$, if $q$=1,2, and $b_p^q(E^k, 1) = -b_p^{-q}(E^k, 2)$, if $q$=4,5.

TABLE V. Parameters of electron-deformation interaction (cm$^{-1}$) in the PrAl$_3$(BO$_3$)$_4$ crystal.

| p | q | $b_p^q(A^1)$ | $b_p^q(A^2)$ | p | q | $b_p^q(E^1,1)$ | $b_p^q(E^2,1)$ |
|---|---|---|---|---|---|---|---|
| 2 | 0 | -262 | 1241 | 2 | 1 | -4323 | -4986 |
| 4 | 0 | 380 | -87 | 4 | 1 | 204.3 | 1075 |
| 6 | 0 | -74.9 | -95.1 | 6 | 1 | 342.4 | 466 |
| 4 | -3 | -4721 | -23.2 | 6 | 5 | -281.5 | -4389 |
| 6 | -3 | -644 | 838 | 2 | -2 | -1.5 | -2165 |
| 6 | 6 | -538 | -333 | 4 | -2 | -790 | 429 |
| | | | | 6 | -2 | 34.3 | 348 |
| | | | | 4 | -4 | 976.5 | -368 |
| | | | | 6 | -4 | 955.5 | -211 |

Assuming a Lorentzian line shape for individual transitions between the hyperfine components of the sublevels Γ and Γ' of two multiplets, the intensity distribution of magnetic dipole absorption at the transition Γ→Γ' at temperature $T$ (the profile of the hyperfine structure envelope) can be described by the formula

$$I(\Gamma \to \Gamma', E) =$$
$$\sum_{j\in\Gamma}\sum_{k\in\Gamma'}\sum_{\alpha=x,y,z}|<j|M_\alpha|k>|^2 \exp[-(E_k-E_0)/k_BT][(E_k-E_j-E)^2+\Delta^2_{\Gamma\Gamma'}]^{-1}, \quad (13)$$

where $\Delta_{\Gamma\Gamma'}$ is the half-width of the spectral line of the electron-nuclear transition, $E_0$ is the energy of the lower hyperfine sublevel of the electron level Γ, and $M$ is the effective dipole moment (electric or magnetic) of the transition under consideration.

In Fig. 3, the calculated six-component hyperfine structures of the lines corresponding to transitions from the ground singlet to the lower doublets in the $^3H_5$ and $^3H_6$ multiplets are compared with the observed profiles of these lines, which contain two broad overlapping components. The width of the dip at the center of the observed lines is significantly larger than the hyperfine structure intervals. We attribute the strong broadening and formation of the doublet structure to the interaction of the RE ions with the field of random deformations induced by both praseodymium impurity ions and intrinsic lattice point defects (molybdenum and bismuth ions in the aluminum and yttrium positions, respectively).

The distribution function of random deformations induced by point defects in an elastically anisotropic trigonal crystal lattice has the form [60]

$$g(e) = \frac{15\xi v_A}{8\pi^3 \gamma_A^2 \gamma_{E^1}^2 \gamma_{E^2}^2} \{[v_A^2 e(A^1)^2 + e(A^2)^2]/\gamma_A^2$$
$$+ [e(E^1,1)^2 + e(E^1,2)^2]/\gamma_{E^1}^2 + [e(E^2,1)^2 + e(E^2,2)^2]/\gamma_{E^2}^2 + \xi^2\}^{-7/2}, \quad (14)$$

where the parameters $v_\Gamma, \gamma_\Gamma$ are determined by elastic constants and $\xi$ is the distribution width proportional to the defect concentration and defect strength. In the calculations, the elastic constants of the HoAl$_3$(BO$_3$)$_4$ crystal [61] were used.

The spectral envelope of the singlet-doublet transition line is obtained by averaging the sum of the shape functions (13) of the hyperfine structure components over the distribution of random deformations (14):

$$I(\Gamma_2 \to \Gamma_3, E) \sim \int g(\boldsymbol{e}) \sum_m I_0(E - E_m(\boldsymbol{e})) \, d\boldsymbol{e}. \tag{15}$$

Here, $E_m$ are the energies of the hyperfine sublevels of the doublet under consideration, equal to the corresponding eigenvalues of the Hamilton operator (1) for fixed components of the deformation tensor $\boldsymbol{e}$. We neglect the hyperfine structure of the $\Gamma_2$ singlet. As follows from the calculations performed, the dominant role in the splitting of the sextet hyperfine structure of electronic doublets into two triplets with an unresolved structure is played by shear deformations, transforming according to the irreducible representation $E^2$. The calculated intensity distributions (15) of absorption lines at singlet-doublet transitions shown in Fig. 2 reproduce well the measurement data when using the width of the deformation distribution function $\xi = 7 \cdot 10^{-5}$ and the half-widths of the shape functions of electron-nuclear transitions $\Delta_{\Gamma\Gamma'} = 0.053$ cm$^{-1}$.

The obtained value of $\xi$ is significantly larger than $\xi \sim 10^{-6}$ found earlier from the simulation of lines' shape of singlet-doublet transitions in the spectra of LiLuF$_4$:Pr$^{3+}$(0.1 at. %) [60] and $R$MO$_4$:Tm$^{3+}$ ($R$ = Y, Lu, $M$ = P, V; Tm concentrations 0.2 and 0.88 at. %) [46]. We note that a deformation splitting of 1.3 cm$^{-1}$ for the first excited doublet in the ground multiplet of Tm$^{3+}$ ions was observed in the THz spectra of TmAl$_3$(BO$_3$)$_4$ [62]. In this crystal grown by the solution-melt technique, Tm$^{3+}$ ions are not an impurity but a part of a regular crystal structure, and the main defects are flux components entering the crystal during the growth process. The large width of the random deformation distribution function obtained from modelling the observed doublet deformation structure of singlet-doublet transitions in the spectra of YAB:Pr$^{3+}$ indicates the possibility of strong correlations and attraction between intrinsic structural defects and praseodymium impurity ions in YAB:Pr$^{3+}$ samples grown by the solution-melt method.

## V. CONCLUSIONS

Transmission spectra of YAl$_3$(BO$_3$)$_4$ single crystals activated with Pr$^{3+}$ ions at concentrations of 1 and 2.5 at.% were measured, including in a magnetic field parallel to the $c$ axis of the crystal. The spectra were recorded in a wide spectral range (2000–23,000 cm$^{-1}$) using high-resolution (up to 0.05 cm$^{-1}$) Fourier spectroscopy, which ensures high accuracy of the wave number scale. Pr$^{3+}$ ions replace Y$^{3+}$ ions in positions with point symmetry group $D_3$. As a result of analysing temperature-dependent (5–300 K) spectra in π- and σ-polarised light, a diagram of CF levels of the Pr$^{3+}$ ion in YAB:Pr$^{3+}$ was constructed and irreducible representations of the point group $D_3$ ($\Gamma_1$, $\Gamma_2$, or $\Gamma_3$) were determined, according to which the wave functions of CF levels are transformed [41]. Measurements in a magnetic field allowed us to determine the $g$ factors of several $\Gamma_3$ doublets. The spectra did not reveal any hyperfine structure caused by the interaction of electrons with the magnetic moment of the nucleus of the only stable isotope of praseodymium, $^{141}$Pr, with nuclear spin $I = 5/2$. However, for a number of singlet-doublet transition lines, a characteristic doublet structure was observed, caused by the action of random lattice deformations. Based on spectroscopic data, calculations were performed using crystal-field theory and the exchange-charge model. A physically justified set of crystal-field parameters was obtained. A calculation of the six-component hyperfine structure of $\Gamma_3$ doublets was performed. The calculated intervals of the hyperfine structure of doublets are significantly smaller than the deformation splitting observed in the spectra. The values of the electron-deformation interaction parameters were calculated within the framework of the exchange charge model. Using the previously obtained distribution function of random deformations induced by point defects in an elastically anisotropic trigonal crystal lattice, the profiles of deformation doublets observed in the optical spectrum were simulated. An estimate of the width of the random deformation distribution function has been obtained. Its large value may be due to a high concentration of impurities – flux components that enter the crystal during solution-melt growth.

## ACKNOWLEDGEMENTS

The authors express their gratitude to N.Yu. Boldyrev for assistance in carrying out some of the measurements. The work was financially supported by the research projects FFUU-2024-0004 and FFUU-2025-0004 of the Institute of Spectroscopy of the Russian Academy of Sciences.

## DATA AVAILABILITY

The data are available from the authors upon reasonable request.